\documentclass[12pt]{article}
 \pdfoutput=1
\usepackage[table]{xcolor}
\usepackage{amsmath,amssymb,exscale}
\usepackage{graphicx}
\usepackage{epsfig}
\usepackage{multicol}
\usepackage{color}
\usepackage{mathrsfs}
\usepackage{blindtext}
 \usepackage{fancyhdr}
\usepackage{hyperref}
\usepackage{cite}
\usepackage{mathtools}
\usepackage{amsmath}
\usepackage{rotating,slashed,amsmath,charter,xcolor,catchfilebetweentags,ifluatex}

\usepackage{graphicx}
\usepackage{sidecap}

\usepackage[latin1]{inputenc} 
\usepackage{multicol}  
 
 \usepackage{titlesec}
 
 \usepackage{rotating,slashed,xcolor,amsfonts,expdlist,charter}

\numberwithin{equation}{section}

\usepackage{xcolor}
\usepackage{sectsty}

\usepackage{mdframed}
\usepackage{titletoc}

\definecolor{secnum}{RGB}{13,151,225}
\definecolor{ptcbackground}{RGB}{212,237,252}
\definecolor{ptctitle}{RGB}{0,177,235}

\titlecontents{lsection}
  [5.8em]{\sffamily}
  {\color{secnum}\contentslabel{2.3em}\normalcolor}{}
  {\titlerule*[1000pc]{.}\contentspage\\\hspace*{-5.8em}\vspace*{5pt}%
    \color{white}\rule{\dimexpr\textwidth-15.5pt\relax}{1pt}}

\usepackage{hyperref}
\hypersetup{colorlinks,bookmarksopen,bookmarksnumbered,citecolor=blus,
linkcolor=redy,pdfstartview=FitH,urlcolor=blus}
\usepackage{slashed}

\definecolor{blus}{cmyk}{1,0.9,0,0.1}
\definecolor{verdes}{cmyk}{0.99,0,0.59,0.65}
\definecolor{rossos}{cmyk}{0,1,1,0.55}
\definecolor{redy}{cmyk}{0,1,1,0.7}
\definecolor{greeny}{cmyk}{0.99,0,0.59,0.98}
\definecolor{green-go}{cmyk}{0.79,0,0.59,0.5}

\usepackage{titlesec}

\def\Lag{\mathscr{L}}

\newcommand{\gs}{f_0}
\newcommand{\gt}{f_2}

\newcommand{\beq}{\begin{equation}}
\newcommand{\eeq}{\end{equation}}

\def\hhref#1{\href{http://arxiv.org/abs/#1}{arXiv:#1}} 

 \def\Lag{\mathscr{L}}
 
\newcommand{\tmtextbf}[1]{{\bfseries{#1}}}
\newcommand{\tmtextrm}[1]{{\rmfamily{#1}}}
\def\bp{M_{P}}

\def\be{\begin{equation}}
\def\ee{\end{equation}}
\def\ba{\begin{array} }

\def\bac{\begin{array} {c}}
\def\bacc{\begin{array} {cc}}
\def\baccc{\begin{array} {ccc}}
\def\bacccc{\begin{array} {cccc}}
\def\ea{\end{array}}
\def\bea{\begin{eqnarray}}
\def\eea{\end{eqnarray}}

\definecolor{red}{rgb}{1,0,0}

\def\psl{\hbox{\hbox{${p}$}}\kern-1.9mm{\hbox{${/}$}}}
\def\dsl{\hbox{\hbox{${\partial}$}}\kern-2.2mm{\hbox{${/}$}}}
\def\Dsl{\hbox{\hbox{${D}$}}\kern-2.6mm{\hbox{${/}$}}}

\def\Lag{\mathscr{L}}

\newcommand{\gappeq}{{\rlap{{\raise}.5ex\text{\ensuremath{>}}}{{\lower}.5ex\text{\ensuremath{\sim}}}}}
\newcommand{\lappeq}{{\rlap{{\raise}.5ex\text{\ensuremath{<}}}{{\lower}.5ex\text{\ensuremath{\sim}}}}}
\newcommand{\I}{\tmtextrm{1{\kern}-.24em l}}

\begin{document}
\topmargin -1.0cm
\oddsidemargin 0.9cm
\evensidemargin -0.5cm

{\vspace{-1cm}}
\begin{center}

\vspace{-1cm}


 {\huge \tmtextbf{ 
\color{rossos} BICEP/Keck data and Quadratic Gravity}} {\vspace{.5cm}}\\

\vspace{1.9cm}

{\large  {\bf Alberto Salvio }
{\em  

\vspace{.4cm}

 Physics Department, University of Rome Tor Vergata, \\ 
via della Ricerca Scientifica, I-00133 Rome, Italy\\

\vspace{0.6cm}

I. N. F. N. -  Rome Tor Vergata,\\
via della Ricerca Scientifica, I-00133 Rome, Italy\\ 

\vspace{0.4cm}

\vspace{0.2cm}

 \vspace{0.5cm}
}

\vspace{0.2cm}

}
\vspace{0.cm}

%
%
 %
%
%

\end{center}

%
 
\noindent --------------------------------------------------------------------------------------------------------------------------------

\begin{center}
{\bf \large Abstract}
\end{center}

\noindent The recent results of the BICEP and Keck collaborations have put stringent bounds on many inflationary models, including some  well-motivated ones. This is certainly the case when gravity remains described by Einstein's theory up to the inflationary scale, but can be avoided by introducing quadratic-in-curvature terms that are effective at that scale.  Recently it has also been shown that these terms can UV complete gravity respecting stability and unitarity. Here the predictions of such quadratic gravity are computed and compared with the BICEP/Keck constraints by focusing on some of the inflationary scenarios that are best-motivated from the particle physics point of view and  are already ruled out in Einstein gravity: (critical) Higgs inflation and natural inflation. The first scenario can be considered as the most economical option as  the inflaton is identified with the only known elementary scalar field in the Standard Model and  the near criticality of the Standard Model is used to remain in the perturbative regime. In the second one a pseudo-Nambu-Goldstone boson contributes to the inflationary dynamics and its potential is naturally flat. It is shown that in both scenarios one can restore the agreement with the observational constraints in quadratic gravity.

  \vspace{0.4cm}

\noindent --------------------------------------------------------------------------------------------------------------------------------

\vspace{1.1cm}

\vspace{2cm}




\tableofcontents

\vspace{0.5cm}


\section{Introduction}\label{Introduction}

The new data released by the BICEP and Keck collaboration~\cite{BICEP:2021xfz} (hereafter BK18) have provided strong new constraints on the tensor-to-scalar ratio $r$ and the spectral index of scalar perturbations $n_s$. In so doing they  have ruled out some of the best-motivated inflationary scenarios.  

Probably the most economical inflationary model is Higgs inflation (HI)~\cite{Bezrukov:2007ep}, where the inflaton is identified with the only elementary scalar of the Standard Model (SM). In its original form, HI agrees well with the observational data, but presents some issues. One is due to the onset of a high energy scale (much below the Planck scale) above which the theory becomes non perturbative~\cite{crit}. Another problem is the necessity of tuning the values of some high-scale parameters~\cite{Salvio:2015kka}. Both these issues can be avoided in critical Higgs inflation (CHI)~\cite{Hamada:2014iga,Bezrukov:2014bra,Hamada:2014wna,Salvio:2017oyf}: a version of HI that occurs when the parameters are chosen in a way that the electroweak (EW) vacuum is right at the border between absolute stability and metastability\footnote{The EW vacuum is called metastable when its lifetime exceeds the age of the universe.}. 

 Another well-motivated inflationary scenario is natural inflation~\cite{Freese:1990rb} where the inflaton is identified with a pseudo-Nambu-Goldstone boson (PNGB) associated with a spontaneously broken approximate global symmetry. Indeed, in this case the quasi-flatness of the inflaton potential is protected by Goldstone's theorem.
 
 Both CHI and natural inflation have thus strong motivations from the particle physics point of view. However, they have been ruled out by the BK18 constraints on $n_s$ and, especially, on $r$, at least if Einstein gravity holds up to the inflationary scales. 
 
 The main purpose of this paper is to show that these (and presumably many more) inflationary scenarios are still perfectly viable if combined with a theory where the gravitational interaction becomes weaker above a certain energy scale (softened gravity)~\cite{Giudice:2014tma,Salvio:2016vxi}, if this energy scale is smaller than the typical inflationary scales. Indeed, in this case the predicted value of $r$ is expected to be drastically reduced. Softened gravity, of course, has to approach Einstein's gravity at small enough energies in a way that all observational constraints are satisfied.
 
  An explicit and viable example of softened gravity is quadratic gravity, where all possible quadratic-in-curvature local terms with dimensionless coefficients are added to the Einstein-Hilbert action (see~\cite{Salvio:2018crh,Salvio:2020axm} for reviews). These terms are unavoidably generated by quantum corrections in {\it any} quantum field theory (QFT)~\cite{Utiyama:1962sn} and make gravity renormalizable~\cite{Stelle:1976gc}.
  
  Some of the extra terms of quadratic gravity were exploited by Starobinsky in his seminal work on inflation~\cite{Starobinsky:1980te}. Starobinsky's model is one of the ones that agree the best with the observational data, predicting 
  \be n_s \approx 1 -2/N\stackrel{N\approx 60}{\simeq}0.967\qquad r \approx 12/N^2\stackrel{N\approx 60}{\approx}  0.003,\ee
   where $N$ is the number of e-folds. With the future installation of additional receivers the experimental system, as discussed in BK18, is projected to reach the sensitivity of $0.003$ for $r$, precisely Starobinsky's prediction.  
  
  Besides being an explicit realization of softened gravity, quadratic gravity also provides a UV completion of Einstein gravity that can be unitary~\cite{Anselmi:2017ygm,Salvio:2019wcp} and sufficiently stable to describe the entire history of the universe~\cite{Salvio:2019ewf, Anselmi:2018bra,Salles:2014rua}. A naive quantization of the theory would lead to negative probabilities because of the presence of higher derivative terms, which manifest themselves through a massive spin-2 field with negative classical kinetic energy. However, as shown in Refs.~\cite{Anselmi:2017ygm,Salvio:2019wcp}, this can be avoided by relaxing some fundamental assumptions usually made in the quantum theory. 
  
  Here we study CHI and natural inflation in quadratic gravity (in its more general form) and compare the theoretical predictions with the new observational constraints. Although the approaches of Refs.~\cite{Anselmi:2017ygm} (based on the ``fakeon" prescription for the massive spin-2 field) and \cite{Salvio:2019wcp} appear to be different, they lead, remarkably, to the same inflationary predictions~\cite{Salvio:2017xul,Anselmi:2020lpp}, at least in the slow-roll approximation and for a single inflaton~\cite{Salvio:2020axm}. Here we also extend this analysis to the general case of an arbitrary number of inflatons.

The paper is organized as follows. In Sec.~\ref{Quadratic gravity} we give some generic results regarding quadratic gravity that are then used in the rest of the paper. CHI in quadratic gravity is studied in Sec.~\ref{CHIsec} and natural inflation in quadratic gravity is the topic of Sec.~\ref{naturalinflation}. Finally, the conclusions are given in Sec.~\ref{Conclusions}. 

\section{Quadratic gravity}\label{Quadratic gravity}

In quadratic gravity Einstein's theory is extended  by adding all  possible local   terms quadratic in the curvature, whose coefficients have the dimensionality of non-negative powers of energy. So the full gravitational Lagrangian (density) is\footnote{We use the mostly-plus signature for the metric and our conventions for the Riemann and Ricci tensors and the Ricci scalar are, respectively, $$R_{\mu\nu\,\,\, \sigma}^{\quad \rho} \equiv \partial_{\mu} \Gamma_{\nu \, \sigma}^{\,\rho}- \partial_{\nu} \Gamma_{\mu \, \sigma}^{\,\rho} +  \Gamma_{\mu \, \tau}^{\,\rho}\Gamma_{\nu \,\sigma }^{\,\tau}- \Gamma_{\nu \, \tau}^{\,\rho}\Gamma_{\mu \,\sigma }^{\,\tau}, \quad R_{\mu\nu} \equiv R_{\rho\mu\,\,\, \nu}^{\quad \rho},\quad R\equiv g^{\mu\nu}R_{\mu\nu}.$$}~\cite{Salvio:2018crh,Salvio:2020axm} 
\be \mathscr{L}_{\rm gravity} =\frac{\bp^2}{2} R - \Lambda_{\rm cc}+\frac{R^2}{6f_0^2}   - \frac{W^2}{2 f_2^2}-\epsilon_1 G-
\epsilon_2\Box R, \label{Lgravity}\ee
where $f_0^2$, $f_2^2$, $\epsilon_1$ and $\epsilon_2$ are four real parameters, $f_0^2$ and $f_2^2$ are both positive to avoid tachyonic instabilities~\cite{Stelle:1976gc,Salvio:2017qkx,Salvio:2018crh}, $\bp$ is the reduced Planck mass and $\Lambda_{\rm cc}$ is the cosmological constant. Also $W^2$ is the square of the Weyl tensor,
  \be W^2\equiv W_{\mu\nu\rho\sigma}W^{\mu\nu\rho\sigma} = R_{\mu\nu\rho\sigma}R^{\mu\nu\rho\sigma} - 2R_{\mu\nu}R^{\mu\nu} + \frac13 R^2, \ee
  and 
  \beq   G\equiv  R_{\mu\nu\rho\sigma}R^{\mu\nu\rho\sigma} - 4 R_{\mu\nu} R^{\mu\nu} + R^2  =  \frac{1}{4}\epsilon^{\mu\nu\rho\sigma}
\epsilon_{\alpha\beta\gamma\delta}R_{\,\,\,\,\,\, \mu\nu}^{\alpha\beta} R_{\,\,\,\,\,\,\rho\sigma}^{\gamma\delta}= \mbox{divs.}, \label{Gdef}
 \eeq
 with $\epsilon_{\mu\nu\rho\sigma}$ the antisymmetric Levi-Civita tensor and  ``divs" denotes the covariant divergence of some current. Thus, both $\Box R$ and $G$ are total covariant derivatives and, once inserted in the action, give rise to boundary terms and do not contribute to the field equations. As a result, these terms do not contribute to the inflationary paths and to the  perturbations and can be neglected here. The $R^2$ term corresponds to an extra massive spin-0 field of gravitational origin, a.k.a the scalaron (see Sec.~\ref{The effective action}), while the $W^2$ term corresponds to the massive spin-2 field, which turns out to have mass $M_2=f_2\bp/\sqrt{2}$.
 
 The full action of the theory is then
\be  S=\int d^4x\sqrt{-g}\left(\mathscr{L}_{\rm gravity}+\mathscr{L}_{\rm matter}\right), \ee
where $g$ is the determinant of the metric and $\mathscr{L}_{\rm matter}$ represents the matter Lagrangian. In order to keep the theory renormalizable $\mathscr{L}_{\rm matter}$ should contain all possible local terms with coefficients having the dimensionality of non-negative powers of energy.

\section{A minimal possibility: (Critical) Higgs inflation}\label{CHIsec}

It is well-known that in order to realise HI there should be a non-minimal coupling $\xi_H$ between the Higgs doublet ${\cal H}$ and the Ricci scalar~\cite{Bezrukov:2007ep}: we have a term $\xi_H |{\cal H}|^2 R$
in $\mathscr{L}_{\rm matter}$. It is also known that  $\xi_H$ needs to be positive in order to realize successful HI~\cite{Isidori:2007vm,Salvio:2019wcp}.

The idea of CHI is to realize inflation along the Higgs direction for values of the parameters such that the EW vacuum is very close to the border between metastability and absolute stability. This happens when the Higgs  coupling $\lambda_H$, that appears in the potential as $\lambda_H |{\cal H}|^4$, is very close to zero around the Planck scale so that the Higgs potential is very flat even for moderate, ${\cal O}(10)$, values of  $\xi_H$. This avoids the high-scale breaking of perturbation theory that is present in the original version of HI for scales much below $\bp$, due to $\xi_H\gg 1$~\cite{crit}. Since $\lambda_H$ is very small at the scales of interest, we can also consistently realize inflation along the Higgs direction by demanding that all the dimensionless couplings of the other scalars are large compared to $\lambda_H$, such that the potential is much steeper along the other directions. For example, as for the effective scalar corresponding to $R^2$, this requires $f_0^2\gg \lambda_H$ at the inflationary scales.

\subsection{The matter sector}

Note that in order to realize CHI in the SM one needs a top mass $M_t\simeq 171$~GeV~\cite{Buttazzo:2013uya}, which is about $2\sigma$ away from the current central value~\cite{Particle data group (top)}. CHI in the SM also predicts  inflationary observables in tension with the most recent Planck and BICEP/Keck bounds~\cite{BICEP:2021xfz,Ade:2015lrj}, namely~\cite{Salvio:2017oyf}
\be n_s \simeq 0.97, \qquad P_R \simeq 2.2\times 10^{-9},\ee
where $P_R$ is the curvature power spectrum and $N\simeq 60$ was assumed. Similar tensions were also pointed out in~\cite{Masina:2018ejw}.
However, we should remember that the SM certainly has to be extended because of well-established facts (e.g. neutrino oscillations and dark matter).

For these reasons it is more realistic to consider CHI in some economical SM extension that can account for all observations. As a benchmark model we consider here the $a\nu$MSM one~\cite{Salvio:2015cja,Salvio:2018rv,Salvio:2021puw,Salvio:2021kya}, which can account for all the experimentally confirmed signals of beyond-the-SM physics (neutrino oscillations and dark matter) and can solve other important issues of the SM (baryon asymmetry, the strong CP problem\footnote{The other two fine-tuning problems of the SM (the Higgs mass and cosmological constant problems) could be addressed, unlike the strong CP one, with anthropic arguments~\cite{Weinberg:1987dv}.}, the metastability of the EW vacuum and inflation) {\it at the same time}.

In the $a\nu$MSM one adds to the SM fields three sterile neutrinos $N_i$ and the  fields of the KSVZ axion model~\cite{Kim:1979if} (two Weyl fermions $q_1$, $q_2$ neutral under ${\rm SU(2)_{\it L}\times U(1)_{\it Y}}$ and a complex scalar $A$).  Correspondingly, one adds to the SM matter Lagrangian, $\Lag_{\rm SM}$,  three terms,
 \be \mathscr{L}_{\rm matter} = \Lag_{\rm SM}+\Lag_{N}+  \Lag_{\rm axion}+\Lag_\xi,\label{full-lagrangian} \ee
 that are defined as follows.  $\Lag_{N}$ represents the $N_i$-dependent term:
  \be  i\overline{N}_i \dsl N_i+ \left(\frac12 N_i M_{ij}N_j +  Y_{ij} L_iH N_j + {\rm h.  c.}\right). \ee
We take the Majorana mass matrix $M$ diagonal and real, $M=\mbox{diag}({\cal M}_1, {\cal M}_2, {\cal M}_3),$ without loss of generality, but the Yukawa matrix $Y$ is generic.
  $\Lag_{\rm axion}$ is  the KSVZ term:
  \be \mathscr{L}_{\rm axion} = i\sum_{j=1}^2\overline{q}_j \Dsl \, q_j +|\partial A|^2  -(y q_2A q_1 +h.c.)-\Delta V(H,A),\nonumber \ee
where $\Delta V({\cal H},A)$ is the $A$-dependent piece of the classical potential
\be \Delta V({\cal H},A) \equiv \lambda_A(|A|^2-  f_a^2/2)^2 + \lambda_{HA} (|{\cal H}|^2-v^2)( |A|^2-f_a^2/2),\nonumber \ee
  $v\simeq 174$~GeV is the EW breaking scale and $f_a$ is the axion decay constant.
  The Yukawa coupling  $y$ is chosen real and positive without loss of generality. Finally, 
 \be \mathscr{L}_\xi =\left[\xi_H (|{\cal H}|^2-v^2) + \xi_A (|A|^2-f_a^2/2)\right]R, 
 \label{gravity-Lag} \ee
where $\xi_A$ is the non-minimal couplings of  $A$ to gravity.

The Peccei-Quinn symmetry breaking induced by $\langle A\rangle=f_a/\sqrt{2}$ leads to the extra quark  mass $M_q = y f_a/\sqrt{2}$ and the extra scalar squared mass
\be M_A^2 = f_a^2\left(2\lambda_A +\mathcal{O}\left(\frac{v^2}{f_a^2}\right)\right).  \label{MA1} \ee
Since $f_a\gtrsim 10^8$~GeV (see Ref.~\cite{DiLuzio:2020wdo} for a review), the $\mathcal{O}\left(v^2/f_a^2\right)$ term is tiny and can be neglected.
 
 Let us now discuss the other generic observational bounds that are relevant for our purposes\footnote{See Refs.~\cite{Salvio:2015cja,Salvio:2018rv,Salvio:2021puw} for a  discussion of the remaining observational bounds.}. Regarding the active neutrinos, here we take  the  currently most precise values reported in~\cite{Esteban:2020cvm,deSalas:2020pgw} for normal ordering (which is currently preferred) of the following quantities: $\Delta m^2_{21}$, $\Delta m^2_{31}$ (where $\Delta m_{ij}^2 \equiv m_i^2-m_j^2$ and $m_i$ are the active-neutrino masses), the active-neutrino mixing angles  and the CP phase in the Pontecorvo-Maki-Nakagawa-Sakata (PMNS) matrix. One must also take into account the cosmological upper bound on the sum of the neutrino masses~\cite{Aghanim:2018eyx}
 \be \sum_i m_i <0.12~\mbox{eV}. \ee 
 
 Regarding the SM sector, we also have to fix the values of the relevant SM couplings at the EW scale, which we identify conventionally with the top mass $M_t\simeq 172.5$ GeV~\cite{Particle data group (top)}. In the $a\nu$MSM (unlike in the SM) CHI can be realized even by taking the central value $M_t\simeq 172.5$ GeV for the top mass. We take the values of the relevant SM couplings at the EW scale computed in~\cite{Buttazzo:2013uya}, which gives these quantities in terms of $M_t$, the Higgs mass $M_h\simeq 125.1$ GeV~\cite{Particle data group}, the strong fine-structure constant renormalized at the $Z$ mass, $\alpha_s(M_Z) \simeq 0.1184$~\cite{Bethke:2012jm} and the $W$ mass $M_W\simeq 80.379$ GeV~\cite{Particle data group} (see the quoted references for the uncertainties). 
 
 Regarding the axion, besides the lower bound $f_a\gtrsim 10^8$~GeV we have an upper bound on $f_a$  that is obtained by requiring that dark matter is not overproduced and the initial misalignment angle is not tuned. Such bound has a mild logarithmic dependence on $\lambda_A$ (for e.g.~$\lambda_A$ around $\sim 10^{-1}$, it is about $5\times 10^{10}$~GeV~\cite{Salvio:2021puw}). Dark matter in this model is generically a mixture of the axion and the lightest sterile neutrino, forming a two-component dark matter, although in some specific cases only one of these components can account for the whole dark matter (see Ref.~\cite{Salvio:2021puw} for details).

\subsection{Renormalization-group equations}
\label{RGEs}

To study HI one needs the 
effective Higgs potential. This is constructed following the prescription given in Ref.~\cite{Salvio:2018rv}, which is consistent with quadratic gravity. To do so one has to use the renormalization group equations (RGEs), which can be derived at 1-loop level and including the effects of $f_0$ and $f_2$ from the general formalism of  Ref.~\cite{agravity}. Here we have applied this formalism to derive the RGEs of the $a\nu$MSM coupled to quadratic gravity to obtain the result that we now describe.

For a generic coupling $g$ defined in the $\overline{\rm MS}$ renormalization scheme we write the RGEs as
\be \frac{dg}{d\tau}= \beta_{g},\ee
where $d/d\tau\equiv \bar{\mu}^2\, d/d\bar{\mu}^2$ and $\bar{\mu}$ is the $\overline{\rm MS}$ renormalization energy scale. The $\beta$-functions  $\beta_{g}$ can also be expanded in loops: 
\be  \beta_{g} =  \frac{\beta_{g}^{(1)}}{(4\pi)^2}+ \frac{\beta_{g}^{(2)}}{(4\pi)^4}+ ... \, ,\ee 
where $ \beta_{g}^{(n)}/(4\pi)^{2n}$   is the $n$-loop contribution.

We start from energies much above $M_A$, $M_q$,  ${\cal M}_{i}$, the scalaron mass and $M_2$. In this case, the 1-loop RGEs of all relevant couplings  are found to be 
{\allowdisplaybreaks\bea  \beta_{g_1^2}^{(1)}& =&    \frac{41g_1^4}{10}, \qquad   \beta_{g_2^2}^{(1)} =- \frac{19g_2^4}{6},\qquad\beta_{g_3^2}^{(1)}  = -\frac{19 g_3^4}{3},\nonumber\\   \beta_{y_t^2}^{(1)}  & =& y_t^2\left(\frac92 y_t^2-8g_3^2-\frac{9g_2^2}{4}-\frac{17g_1^2}{20}+\frac{15}{8}f_2^2 + {\rm Tr}(Y^\dagger Y )\right),\nonumber\\ 
  \beta_{\lambda_H}^{(1)} & =&\left[12\lambda_H+6y_t^2-\frac{9g_1^2}{10}-\frac{9g_2^2}{2}+\frac52 \gt^2+\frac12 \gs^2 (1+6\xi_H)^2+2\, {\rm Tr}(Y^\dagger Y)\right]\lambda_H \nonumber\\ &&\hspace{-0.7cm}-\, 3y_t^4 +\frac{9 g_2^4}{16}+\frac{27 g_1^4}{400}+\frac{9 g_2^2 g_1^2}{40}+\frac{\lambda_{HA}^2}{2}+\frac{\xi_H^2}{4}\left[5 \gt^4+\gs^4(1+6\xi_H)^2\right] - {\rm Tr}((Y^\dagger Y)^2), \nonumber\\ 
 \beta_{\lambda_{HA}}^{(1)} & =& \left(3y_t^2-\frac{9g_1^2}{20}-\frac{9g_2^2}{4}+6\lambda_H+4\lambda_A +\, {\rm Tr}(Y^\dagger Y ) + 3y^2+2\lambda_{HA}\right. \nonumber \\
 &&\left. +\frac52 \gt^2+\frac{\gs^2}{12}\left[(6\xi_A+1)^2+(6\xi_H+1)^2+4(6\xi_A+1)(6\xi_H+1)\right] \right) \lambda_{HA}\nonumber \\
  && -\xi_H\xi_A\left[ \frac52\gt^4  + \frac12\gs^4(6\xi_A+1)(6\xi_H+1)\right], \nonumber\\ 
 \beta_{\lambda_A}^{(1)} & =& \lambda_{HA}^2+10\lambda_A^2+6y^2 \lambda_A- 3 y^4+ \frac{\xi_A^2}{4} \left[5 \gt^4+\gs^4 (1+6\xi_A)^2\right] + \frac{\lambda_A}{2} \left[5 \gt^2+\gs^2(1+6\xi_A)^2\right],\nonumber\\ 
   \beta_{Y}^{(1)} & =&Y  \left[\frac32 y_t^2-\frac{9}{40} g_1^2-\frac98 g_2^2+\frac34 Y^\dagger Y+\frac12 {\rm Tr}(Y^\dagger Y )+\frac{15}{8}f_2^2\right],\nonumber \\ 
 \beta_{y^2}^{(1)} & =&y^2\left(4y^2-8 g_3^2+\frac{15}{8}f_2^2\right),\nonumber \\ 
  \beta_{\xi_H}^{(1)} &=& (1+6\xi_H)\left(\frac{y_t^2}{2}+\frac{{\rm Tr}(Y^\dagger Y )}6 -\frac{3g_2^2}8  - \frac{3 g_1^2}{40}+\lambda_H\right) -\frac{\lambda_{HA}}{6}(1+6\xi_A)\nonumber \\
&&+\frac{\gs^2}{6}\xi_H(1+6\xi_H)(2+3\xi_H) - \frac56 \frac{\gt^4}{\gs^2}\xi_H, \nonumber \\
 \beta_{\xi_A}^{(1)}  &=& (1+6\xi_A) \left(\frac{y^2}{2}+\frac{2}{3}\lambda_A\right)
-\frac{\lambda_{HA}}{3} (1+6\xi_H)+\frac{\gs^2}{6}\xi_A(1+6\xi_A)(2+3\xi_A) - \frac56 \frac{\gt^4}{\gs^2}\xi_A, \nonumber \\
 \beta_{f_2^2}^{(1)}  &=&-\gt^4\bigg(\frac{133}{20} +\frac{N_V}{10}+\frac{N_f}{40}+\frac{N_s}{120}
\bigg), \nonumber \\
\beta_{f_0^2}^{(1)}  &=& \frac56 \gt^4 + \frac52 \gt^2 \gs^2 + \frac{5}{12} \gs^4 +\frac{\gs^4}{12}\left[2 (1+6\xi_H)^2+(1+6\xi_A)^2\right] \label{f0RGE}
\eea}
where  $g_3$,  $g_2$ and  $g_1=\sqrt{5/3}g_Y$ are the gauge couplings of SM gauge group ${\rm SU(3)}_c $, ${\rm SU(2)_{\it L}}$ and   ${\rm U(1)_{\it Y}}$, respectively, $y_t$ is the top Yukawa coupling.  Also $N_V$, $N_f$, $N_s$ are the number of vectors, Weyl fermions and real scalars.
In the SM $N_V=12$, $N_f = 45$, $N_s = 4$, while in the $a\nu$MSM $N_V=12$, $N_f = 54$, $N_s = 6$.

Since the SM couplings evolve in the full range from the EW to the Planck scale it is appropriate to use for them the 2-loop RGEs\footnote{In the absence of gravity the RGEs for a generic quantum field theory  were computed up to 2-loop order in~\cite{MV}.}, $\beta_{g}^{(2)}$, which can be found in~\cite{Salvio:2021puw}. The matching at the mass thresholds due to the new scalar $A$ and fermions $N_i$, $q_1$ and $q_2$ is performed as explained in Ref.~\cite{Salvio:2018rv} and for the mass threshold of the scalaron and the massive spin-2 field we follow Ref.~\cite{agravity}.
 
 \subsection{(Critical) Higgs inflation in quadratic gravity}\label{CHIsecs}
 Once the Higgs effective potential is constructed one can study inflation taking into account the quadratic-in-curvature terms that can affect the inflationary paths as well as the perturbations~\cite{Salvio:2017xul,Anselmi:2020lpp}. The main difference one obtains in the slow-roll approximation (which remains valid here) is a suppression of the tensor-to-scalar ratio $r$ compared to the value $r_E$ in Einstein's theory:
 \be r=\frac{r_E}{1+2H^2/M_2^2}, \label{rg}\ee
 where $H$ represents here the Hubble rate at the inflationary scale. This is a dynamical quantity that needs to be calculated from the effective potential by requiring an appropriate number of e-folds $N$, e.g.~$N=60$, and to satisfy all observational bounds. The spectral index of scalar perturbations $n_s$ remains instead the same as in Einstein gravity in the slow-roll approximation.
 
Before going to the comparison with the observational data some remarks are in order.
 First, as shown in~\cite{Salvio:2017xul}, the relation in~(\ref{rg}) holds for any slow-roll inflationary models, even in the presence of a generic number of inflatons.

Another remark is that, as discussed below, the formula in~(\ref{rg}) is also valid in the effective field theory, where the $W^2$ term is  a small perturbation and no additional gravitational degrees of freedom are present besides the massless graviton (henceforth the EFT). In order for the EFT to be in its regime of validity, however,  $H$ has to be small enough,  at least one order of magnitude below $M_2$.  Indeed, $M_2$ is the EFT cutoff, while $H$ is the typical  energy in the gravity sector during inflation. 
These  statements can be understood as follows.\begin{itemize}
  \item According to Einstein equations, gravity is sourced by the ratio $T_{\mu\nu}/\bp^2$ of the energy momentum tensor $T_{\mu\nu}$ and $\bp^2$. 
  \item So the typical gravitational energy during inflation is $\sqrt{V_I/\bp^2}$, where $V_I$ is the corresponding energy density (stored in the inflaton sector). Using again the Einstein equations one finds that $\sqrt{V_I/\bp^2}$ is also of order $H$.
  \item Thus the fact that $H$ is well below $M_2$ guarantees that the massive spin-2 field is not excited and one can use the above-mentioned EFT. This condition  for the validity of such EFT can also be checked directly: from the results in~\cite{Salvio:2017xul} we know that the modes corresponding to the massive spin-2 field decouple when $H$ is below $M_2$. 
  \end{itemize}
  Eq.~(\ref{rg}), however, is also valid in the EFT because this is an effect of the $W^2$ term on the massless graviton mode, which is {\it not} integrated out and explicitly appears in the infrared.  Whether  $r$ is small enough to make the theoretical predictions compatible with the observations may depend on the specific inflationary model.
 
\begin{figure}[t]
\begin{center}
 \includegraphics[scale=0.8]{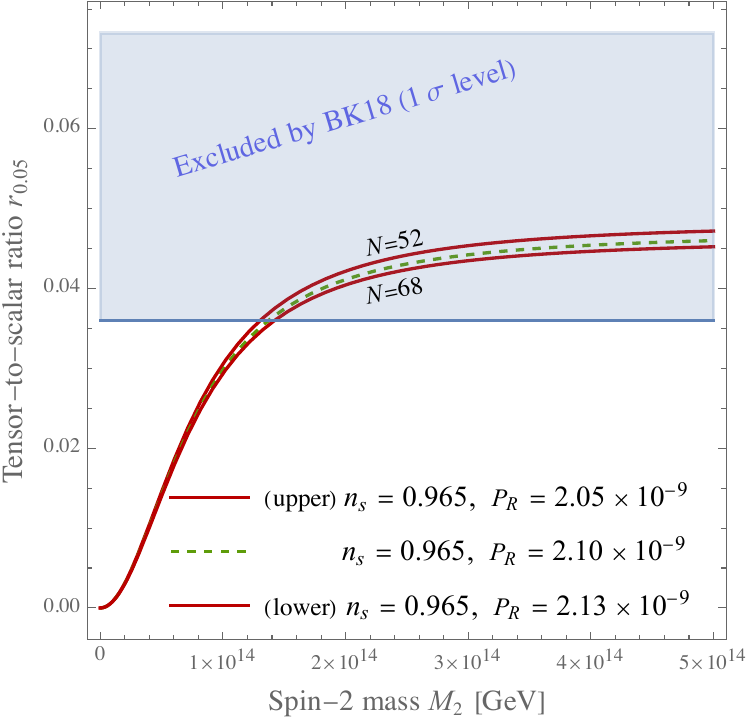}

 \end{center}
   \caption{\em The observable predictions of critical Higgs inflation in quadratic gravity compared to the BICEP/Keck constraints (BK18). We chose  the input values  ${\cal M}_1 = 10^{11}\,{\rm GeV}$, ${\cal M}_2 = 6.4	 \times 10^{13}\,{\rm GeV}$, ${\cal M}_3 > \bp$, 
    $f_a \simeq 3.5 \times 10^{10}\,{\rm GeV}$,
     $\lambda_A(M_A) \simeq 0.1$, $\lambda_{HA}(M_A) \simeq 0.016$, $y(M_A)\simeq  0.1$, $\xi_H(M_A)\simeq 14$, $\xi_A(M_A)\simeq -2.6$,  $f_0\sim 1$,
    $M_2=\bp f_2(M_A)/\sqrt{2}$ and the Casas-Ibarra method~\cite{Casas:2001sr} 
    used in~\cite{Salvio:2018rv} has been adopted. The white region is allowed by BK18, while the shaded one is excluded at 95\% C.L..}
\label{CHI}
\end{figure}

 In Fig.~\ref{CHI} we compare the predictions for the main inflationary observables given by CHI with the BK18 constraints. We find that in Fig.~\ref{CHI}  the contribution of $f_2$ and $f_0$ to the RGEs  give small corrections in the plot. It turns out that the dominant effect of the gravitational couplings is to modify $r$ according to Eq.~(\ref{rg}). $f_0$ has been chosen of order 1 to have a steep scalaron direction in the Einstein frame potential so that (critical) Higgs inflation can be realized. This setup is naturally consistent with the RGEs for $\xi_H\sim{\cal O}(10)$ (see in particular the RGE of $f_0^2$ in~(\ref{f0RGE})).

 It is important to remark that for the parameter values used in Fig.~\ref{CHI} the $a\nu$MSM in quadratic gravity does not only give a successful implementation of CHI, but also accounts for neutrino oscillations, dark matter and baryon asymmetry, solves the strong CP problem, stabilizes the EW vacuum (all at the same time) and can fit well the EW observables. This is the reason why CHI has been implemented here in the $a\nu$MSM rather than in the SM.
 
 The corresponding value of the inflationary Hubble rate is nearly constant with respect to $N$ around $N\simeq 60$,
 \be H \simeq  \left[5.376  - 0.001 (N - 60)\right]10^{13}~{\rm GeV}\simeq \left[2.2079  - 0.0005 (N - 60)\right]10^{-5}\bp. \label{H-CHI}\ee
These values of $H$ are consistent with  the observational bound from Planck~\cite{Ade:2015lrj}
\be H< 2.7\times 10^{-5}\bp. \quad  (2\sigma~\mbox{level})
 \label{PlanckHboound}\ee
 Having determined $H$ one can now compute the predictions for $r$ through Eq.~(\ref{rg}) and by using standard slow-roll formul\ae.
 
Looking at Fig.~\ref{CHI} and Eq.~(\ref{H-CHI}) we first note that, within the above-mentioned EFT, the fact that $H$ should be well below $M_2$  precludes the compatibility between the theoretical prediction of $r$ and the observations in the CHI case.
On the other hand, we observe that $r$ starts being allowed by the BK18 constraints when $M_2$ goes below $\sim 10^{14}$~GeV. Note that one can have perfectly allowed values of $r$ even when the consistency condition $H<2M_2$ present in the approach of Ref.~\cite{Anselmi:2020lpp}  is taken into account.  
 It is also remarkable that the predicted value of $n_s$ is steadily about $0.965$ in agreement with the BK18 constraints and the amplitude of the curvature power spectrum $P_R$ agrees with the Planck constraint~\cite{Ade:2015lrj}
 \be P_R=(2.10 \pm 0.03) 10^{-9} \label{PRPlanck}\ee
 for $N$ around 60.


 \section{A natural option: a pseudo-Goldstone boson-driven inflation}\label{naturalinflation}
\subsection{The effective action}\label{The effective action}
In this case the inflaton (or, more generally, one of the scalar fields driving inflation) $\phi$ is not  fundamental, but rather a composite field, which emerges as a PNGB of a spontaneously broken global symmetry $G$~\cite{Freese:1990rb}. Here we assume that this symmetry breaking occurs non perturbatively in a strongly interacting and confining QFT (analogous to quantum chromodynamics).
Being $\phi$ a PNGB, $\phi/f$ is an angular field variable, where $f$ is the spontaneous symmetry breaking scale of the QFT in question. In order for $\phi$ to successfully drive inflation in agreement with observations $f$ has to be around the Planck scale. To obtain a non-trivial potential for $\phi$, which is essential for inflation, one has to introduce explicit $G$-breaking terms in the action, which break in turn the shift symmetry of $\phi$ at some scale: we denote this scale by $\Lambda^2/f$, with $\Lambda$ some auxiliary energy parameter.  In order to reproduce the small observed value of $P_R$ in~(\ref{PRPlanck}) one needs $\Lambda\ll f$.

Let us see now how to combine the idea of natural inflation with quadratic gravity in its most general form\footnote{Natural inflation has been previously studied only in the presence of some quadratic-in-curvature terms~\cite{Salvio:2019wcp,Salvio:2021lka}.}. 
At the energy scales relevant for inflation the generically covariant effective action of $\phi$ and $g_{\mu\nu}$ can be taken to be
\be S_{\rm eff} =\int d^4x\sqrt{-g}\left[\frac{F(\phi)}{2} R+A(\phi)W^2+B(\phi)R^2 +C(\phi) G+D(\phi)\Box R+\frac12 \phi\Box \phi
-V(\phi)\right], \label{Seff}\end{equation}
where 
 $A, B, C, D, F$ and $V$ are functions of $\phi$, which emerge from the shift-symmetry breaking.
The terms involving higher derivatives compared to those present in~(\ref{Seff})  have been neglected because, although $\phi$ can be at the Planck scale or above, the energy scales probed by inflation are much below $\bp$, see~(\ref{PlanckHboound}). Quadratic-in-curvature terms, however, must be included in $S_{\rm eff}$ because they already appear in the starting action, see~(\ref{Lgravity}).

The functions $A, B, C, D$ and $F$ should be equal up to small corrections to the corresponding quantities in the original action, $A(\phi)\simeq -1/(2f_2^2)$, $B(\phi)\simeq 1/(6f_0^2)$, $C\simeq -\epsilon_1$, $D\simeq -
\epsilon_2$  and $F\simeq \bp^2$. This is because such corrections should be of order of the scale $\Lambda^2/f$ at which the shift symmetry   is broken divided by the Planck mass\footnote{One has to divide by $\bp$ because $A, B, C, D$ and $F$ are dimensionless and because such small corrections  should disappear when gravity is decoupled, that is for $\bp\to\infty$.} and so  are of order
\be \frac{\Lambda^2}{f \bp}\ll \Lambda/\bp\ll\frac{f}{\bp}.\ee 
Consequently, $S_{\rm eff}$ can be approximated by 
 \begin{equation} S_{\rm eff} \simeq \int d^4x\sqrt{-g}\left[\frac{\bp^2}{2} R+\frac{R^2}{6f_0^2}-\frac{W^2}{2f_2^2}-\frac12 (\partial \phi)^2-V(\phi)\right].
 \label{SI}\end{equation}

 The potential $V$
 of the PNGB $\phi$ is periodic with period $2\pi f$~\cite{Freese:1990rb}
 \be V(\phi) = \Lambda^4 \left(1+\cos\left(\frac{\phi}{f}\right)\right) +\Lambda_{\rm cc} \label{VInf}\ee
Here we explicitly  see that the scale $\Lambda$  corresponds to breaking of the shift symmetry. The constant $\Lambda_{\rm cc}$ accounts for  
the (tiny and positive) cosmological constant responsible for the observed dark energy and is completely negligible during inflation, which occurs at a much larger energy scale\footnote{A  more general PNGB potential would be 
 \be V(\phi) = \Lambda^4 \left(1\pm\cos\left(\frac{n\phi}{f}\right)\right)+\Lambda_{\rm cc}\ee
  with $n$ being an integer. However, we can go from the $+$ to the $-$ sign above by shifting $\phi\to \pi f/n$ and we can absorb $n$ in a redefinition of $f$, which would then be the spontaneous symmetry breaking of $G$ divided by $n$.  So as far as inflation is concerned we can just take~(\ref{VInf}).}. 
Given that $V$ is even and periodic with period $2\pi f$ we can focus on the field interval 
\be\phi\in[\pi f, 2\pi f].\ee

As well-known, the $R^2$ term  corresponds to an additional scalar $z>0$~\cite{Salvio:2018crh,ainflation}
\begin{equation} S_{\rm eff} = \int d^4x\sqrt{-g}\left[\frac{\bp^2}{2}R-\frac{W^2}{2f_2^2}-\Lag_{\rm kin}-U \right], \label{Gammast}\end{equation}
where 
\be  \Lag_{\rm kin} \equiv \frac{6\bp^2}{z^2}
 \frac{(\partial \phi)^2 + (\partial z)^2}{2},\label{Lkin} \ee
 \be U(\phi,z)\equiv  \frac{36\bp^4}{z^4}\bigg[{V(\phi)}+   \frac{3f_0^2}{8}\left(\frac{z^2}{6} -\bp^2\right)^2 \bigg].\label{Uphiz}\ee
 Notice that for large $f_0^2$, the Einstein-frame potential $U$ forces $z^2=6\bp^2$ and we obtain pure-natural inflation.
The opposite extreme case is when $\Lambda$ is large, in which case $U$ forces $\phi$ to lie at the minimum of its potential, $\phi = \pi f$, and one recovers inflation along the $z$ direction. We can go to the pure-natural and to the pure-$z$ inflation by taking respectively a small and large value of the following parameter  
\be \rho\equiv \frac{\Lambda^2}{\sqrt{6}f_0\bp^2}. \label{rhodef}\ee 
This is also confirmed by the fact that the masses of $\phi$ and $z$ are respectively
\be m_\phi = \frac{\Lambda^2}{f}, \qquad m_z =  \frac{f_0M_P}{\sqrt{2}}, \ee
where  the tiny cosmological constant has been neglected. These masses are computed by diagonalizing the Hessian matrix of $U$ at its global minimum, $\phi=\pi f$ and $z=\sqrt{6}\bp$, which is also the only minimum of $U$ (modulo the $2\pi f$ periodicity).

 \subsection{Covariant multifield formalism and inflationary paths}\label{Covariant multifield formalism and inflationary paths}
  
  We consider a general setup, which has never been explored before, where natural inflation is studied in quadratic gravity with both the $R^2$ and the $W^2$ terms.
   In order to derive the needed inflationary formul\ae~it is convenient to start with an even more general framework. Notice that~(\ref{Gammast}) belongs to the class of multifield inflationary actions
\be S_{\rm eff}=\int d^4x  \sqrt{-g} \,\bigg[ \frac{\bp^2}{2}R-\frac{W^2}{2f_2^2}  - 
\frac{K_{ij}(\Phi) }{2}\partial_\mu \phi^i\partial^\mu \phi^j-
U(\Phi)
\bigg], \label{action}
 \ee
 where $\Phi$ is an array of scalar fields with components $\phi^i$ and $K_{ij}$ is a field metric\footnote{In our case $K_{ij}$ is proportional to $\delta_{ij}$ with proportionality factor equal to $6\bp^2/z^2$ (see Eq.~(\ref{Lkin}))}.
  For a generic function $\mathscr{F}$ of $\Phi$, we define $\mathscr{F}_{,i}\equiv \partial \mathscr{F}/\partial \phi^i$. Moreover, $\gamma^i_{jk}$ is the Levi-Civita connection in the scalar-field space
 \be \gamma^i_{jk}\equiv \frac{K^{il}}{2}\left(K_{lj,k}+K_{lk,j}-K_{jk,l}\right) \ee
 and $K^{ij}$ is the inverse of the field metric $K_{ij}$ (which is used to raise and lower the scalar indices $i,j,k, ...$); e.g. $\mathscr{F}^{,i}\equiv K^{ij}\mathscr{F}_{,j}$. The connection $\gamma^i_{jk}$ allows to define a covariant derivative $D_j$ in the field space: for a vector $\Phi$-dependent field  $\mathscr{V}_i$  
 \be D_j\mathscr{V}_i\equiv \mathscr{V}_{i;j} \equiv\mathscr{V}_{i,j} - \gamma^k_{ij}\mathscr{V}_k. \ee
 
    Let us consider now the Friedmann-Robertson-Walker (FRW) metric
 \be ds^2 = dt^2 -a(t)^2 \left[dr^2 +r^2(d\theta^2 +\sin^2\theta d\phi^2)\right],  \ee 
During inflation the energy density is dominated by the scalar fields so the curvature contribution has been neglected.
 Note that the FRW metric is conformally-flat and, therefore, the inflationary paths are not affected by the $W^2$ term. As we will discuss in Sec.~\ref{Perturbations}, the $W^2$ term  modifies instead the perturbations.
The Einstein equations and the scalar equations imply the following equations for $a(t)$ and the spatially homogeneous fields $\phi^i(t)$  
  \bea 
   &&H^2=\frac{K_{ij} \dot\phi^i\dot \phi^j/2+U}{3 \bar M_{\rm Pl}^2},\label{EOM1}\\ &&\ddot \phi^i +\gamma^i_{jk}\dot \phi^j\dot \phi^k +3H\dot \phi^i+U^{,i}  =0,   \label{ExactSE} \eea
 where a dot denotes a derivative with respect to $t$ and  $H\equiv \dot a/a$. For a generic vector ($\Phi$-dependent) field  $\mathscr{V}^i$ one also introduces another vector field
\be {\cal D}_t \mathscr{V}^i \equiv  \dot\phi^j D_j\mathscr{V}^i \equiv \mathscr{\dot V}^i +\gamma^i_{jk}\mathscr{V}^j\dot\phi^k,\ee
which allows to write~(\ref{ExactSE}) in the more compact form
${\cal D}_t \dot \phi^i+3H\dot \phi^i+U^{,i}=0.$

Inflation in general takes place when 
\be \epsilon \equiv -\frac{\dot H}{H^2}  < 1. \ee 
We are interested in an expanding universe, $H>0$, so~(\ref{EOM1}) leads to 
\be H(\phi,\dot \phi) = \sqrt{\frac{K_{ij}(\phi) \dot\phi^i\dot \phi^j/2+U(\Phi)}{3 \bar M_{\rm Pl}^2}}, \label{Hfromphi}\ee
which provides, combined with Eq.~(\ref{ExactSE}), an equation for the scalars $\phi^i$ only. Once a solution $\phi^i_0$ of the latter equation is obtained one can compute $H$ by using~(\ref{Hfromphi}). Having determined $H$  the number of e-folds $N$ (from a generic time $t$ until the time $t_e$ when inflation ends) is given by
 \be N = \int^t_{t_e} dt' H. \label{Ndef} \ee 
 
Now, it is convenient to introduce the unit vector $\hat\sigma^i$ tangent to the inflationary path $\phi_0^i$,
\be \hat\sigma^i\equiv \frac{\dot\phi_0^i}{\dot\sigma}, \qquad \dot\sigma\equiv \sqrt{K_{ij}(\Phi_0)\dot\phi_0^i\dot\phi_0^j}. \ee
Besides $\hat\sigma^i$ it is also useful to introduce the set of unit vectors $\hat s^i$ orthogonal to the inflationary path. In the presence of two inflatons
  we have only one of such orthogonal unit vectors   (see e.g.~\cite{Gundhi:2018wyz}) that is proportional to the acceleration vector $\omega^i \equiv  {\cal D}_t \hat\sigma^i$
  and, for actions of the form~(\ref{Gammast}), its explicit expression is
 \be  \hat s^1 \equiv  \frac{\dot z_0}{\dot\sigma}, \qquad \hat s^2 \equiv  \frac{-\dot \phi_0}{\dot\sigma}.\ee

As well-known (see e.g.~\cite{Chiba:2008rp}), when inflation is driven by more than one scalar field, slow-roll occurs if two conditions are satisfied:
\be \epsilon \simeq  \frac{  \bp^2 U_{,i}U^{,i}}{2U^2} \ll 1. \label{1st-slow-roll}\ee 
\be \left|\eta^{i}_{\,\,\, j}\right|  \ll 1, \quad \mbox{and}\quad  \left|\frac{U^{,i}}{U^{,j}}\right| = {\cal O}(1),\quad \mbox{where}\quad \eta^{i}_{\,\,\, j}\equiv \frac{\bp^2 U^{;i}_{\,\,\, ;j}}{U}, \label{2nd-slow-roll} \ee
where in~(\ref{2nd-slow-roll}) we have to restrict our attention to  $i$ and $j$ with the smallest values of $U^{,i}$ and $U^{,j}$, because the directions in field space with a steeper potential do not correspond to slow-roll inflation. In fact it is easier to project $\eta^{i}_{\,\,\, j}$ along the inflationary path, so condition~(\ref{2nd-slow-roll}) leads to
\be |\eta_{\sigma\sigma}|\ll 1, \qquad \mbox{where} \qquad \eta_{\sigma\sigma}\equiv  \hat\sigma^i\hat\sigma^j \eta_{i j}.\ee
When the slow-roll conditions are satisfied Eqs.~(\ref{ExactSE}) and~(\ref{Hfromphi}) simply read 
\be\dot \phi^i\simeq -\frac{U^{,i}}{3H},\qquad   H \simeq \sqrt{\frac{U(\Phi)}{3  \bar M_{\rm Pl}^2}}. \label{slow-roll-eq}\ee
When slow roll holds $N$ can be considered as a function of the scalar fields $\Phi$ by requiring $t-t_e$ in Eq.~(\ref{Ndef}) to be the time it takes for the system to reach the end of inflation starting with initial condition $\Phi$: this is because in the slow-roll approximation the  scalar field equations are of first order (see the first equation in~(\ref{slow-roll-eq})).

 \subsection{Perturbations}\label{Perturbations}
 
 As we mentioned above, at the level of the perturbations the $W^2$ term has a non-trivial effect.

 Let us start with the curvature perturbation $\mathcal{R}$. In~\cite{Salvio:2017xul} it was shown that $\mathcal{R}$ is the same as in Einstein gravity in the slow-roll approximation and the associated power spectrum is 
 \be P_R(q)=\left(\frac{H}{2\pi}\right)^2 N_{,i}N^{,i}\label{power-spectrum}\ee
(in this section we compute the power spectra  at horizon exit $q=a H$). To compute $n_s$ we can, therefore, use the standard formula obtained without considering the $W^2$ term~\cite{Chiba:2008rp,Sasaki:1995aw}
\be  n_s =1-2\epsilon - \frac{ 2 }{ \bar M_{\rm Pl}^2 N_{,i}N^{,i}}+\frac{2\eta_{ij}N^{,i}N^{,j}}{N_{,k}N^{,k}}.  \label{nsFormula}
\ee

 Similarly, let us show here that all isocurvature modes due to additional 4D scalars besides the inflaton are also the same as in Einstein gravity in the slow-roll approximation. From the results of~\cite{Salvio:2017xul} the 4D scalar perturbations $\varphi^i\equiv \phi^i-\phi_0^i$ are given, in such approximation, by
\be \varphi^i =\varphi^i_{\rm dS} -\frac{\dot \phi^i_0}{H} \Psi_{\rm dS},
\ee
where  $\Psi$ denotes one of the metric perturbations and $\varphi^i_{\rm dS}$ and $\Psi_{\rm dS}$ are the values of, respectively, $\varphi^i$ and $\Psi$ in the pure de Sitter case. The isocurvature modes are obtained by projecting $\varphi^i$ along the unit vectors (in the scalar field space) $\hat s^i$ that are orthogonal to the background trajectory, $\hat s_{i} \dot\phi^i_0 =0$. Therefore, the isocurvature modes $\hat s_{i} \varphi^i$ are independent of the metric perturbations and so are the same as in Einstein gravity in the slow-roll approximation.

\begin{figure}[t]
\begin{center}
 \includegraphics[scale=0.58]{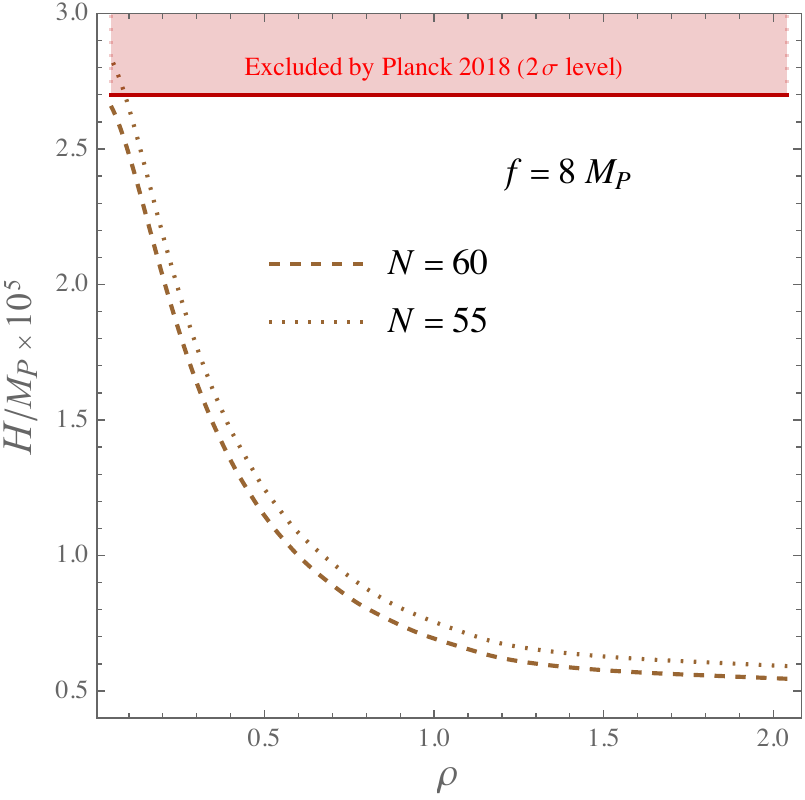}  
  \hspace{1cm}   \includegraphics[scale=0.58]{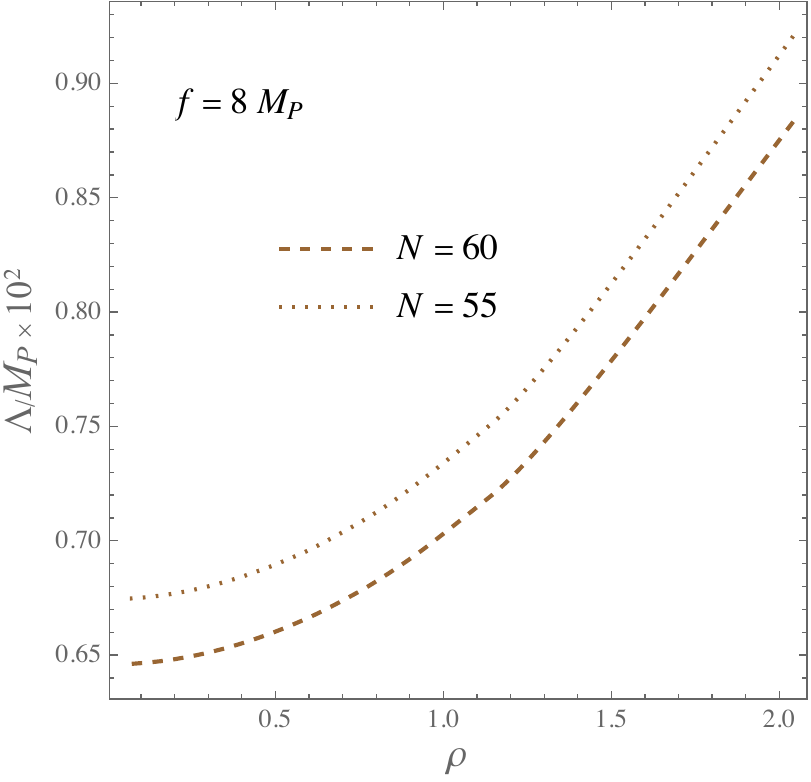} 
 	 
 \end{center}
   
   \caption{\em  {\bf Left plot:} the Hubble rate $H$ at $N$ e-folds before the end of inflation as a function of the $\rho$ parameter defined in~(\ref{rhodef}). {\bf Right plot:} the corresponding value of the parameter $\Lambda$ responsible for the explicit breaking of the global symmetry $G$.}\vspace{0.1cm}
   {\em

  These quantities are obtained by requiring the curvature power spectrum to match the central value in~(\ref{PRPlanck}). Also we set here $f=8\bp$, which reproduces well the observed value of $n_s$.
   }
\label{HLambda}
\end{figure}

\begin{figure}[h!]
\begin{center}
 \includegraphics[scale=0.58]{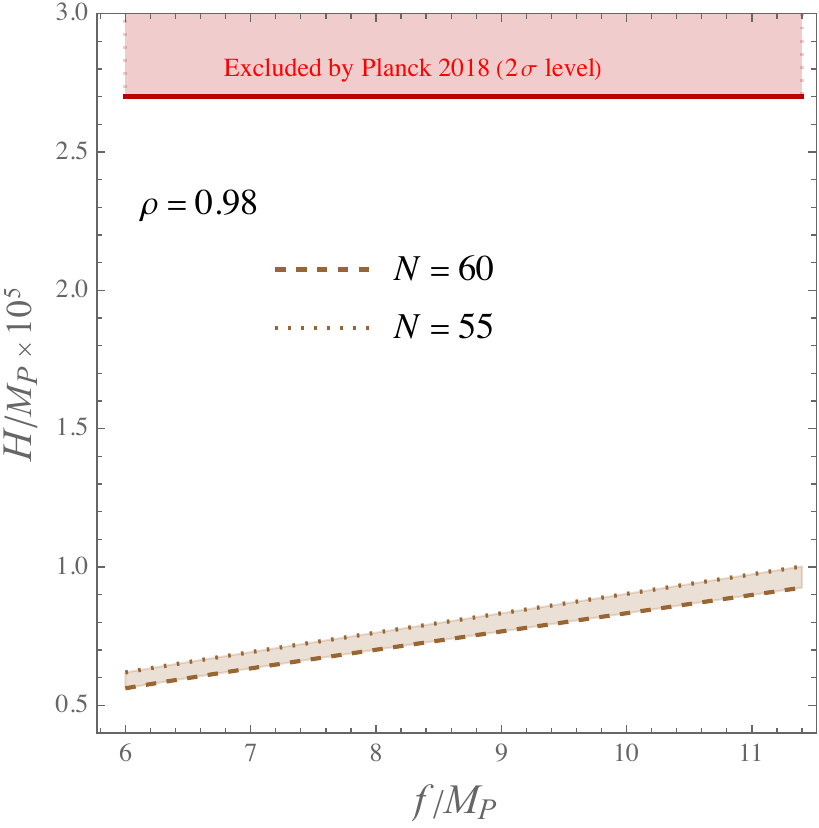}  
  \hspace{1cm}   \includegraphics[scale=0.58]{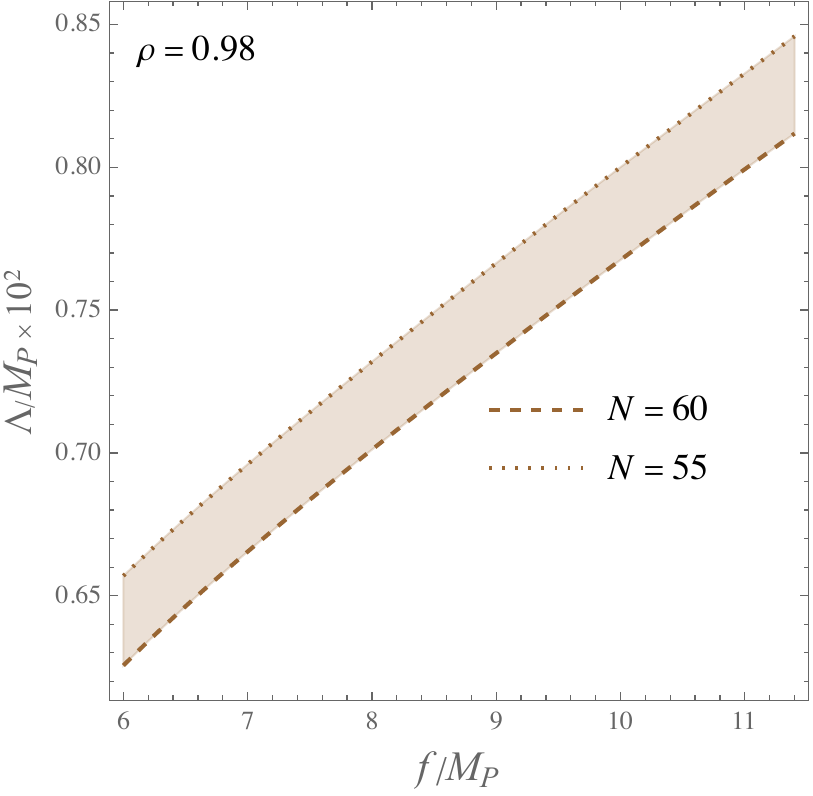} 
 	 
 \end{center}
   
   \caption{\em  {\bf Left plot:} the Hubble rate $H$ at $N$ e-folds before the end of inflation as a function of the  scale $f$. {\bf Right plot:} the corresponding value of the parameter $\Lambda$ responsible for the explicit breaking of the global symmetry $G$.}\vspace{0.1cm}
   {\em

  These quantities are obtained by requiring the curvature power spectrum to match the central value in~(\ref{PRPlanck}). 
   }
\label{HLambdaf}
\end{figure}

Ref.~\cite{Anselmi:2020lpp} only considered one inflaton so it cannot immediately be applied to a multifield inflationary scenario. However, the results above imply that the approach of Ref.~\cite{Anselmi:2020lpp}, which consists in projecting out the massive spin-2 field from the physical spectrum (``fakeon" approach), would give the same predictions for $\mathcal{R}$ and all isocurvature modes due to additional 4D scalars as in Einstein gravity, at least in the slow-roll approximation, which we use here.

Let us move on to the tensor perturbations (the vector perturbations are not observable~\cite{Salvio:2017xul,Anselmi:2020lpp}).
Here is where the $W^2$ term gives its main effect. The corresponding power spectrum is given by~\cite{Salvio:2017xul}
\be  P_t(q) = \frac{1}{1+\frac{2 H^2}{M_2^2}} \frac{8}{\bar M_{\rm Pl}^2} \left(\frac{H}{2\pi}\right)^2. \label{Ptspectrum} \ee 
Also the spectral index $n_t$ of tensor perturbations, defined by $n_t =\frac{d\ln P_t}{d\ln q}$, is~\cite{Salvio:2020axm}
\be n_t = -\frac{2\epsilon}{1+\frac{2 H^2}{M_2^2}} \ee
independently of the number of inflatons.  This result reduces to
\be n_t = -\frac{r}{8} \ee
in single-field inflation (where the Einstein value of the tensor-to-scalar ratio is $16\epsilon$). The spectral indices of scalar and tensor perturbations allows us to compute $r$ at different scales; we denote by $r_{0.002}$ the value of $r$ at the reference momentum scale $0.002~{\rm Mpc}^{-1}$, used by the Planck collaboration~\cite{Ade:2015lrj} and BK18. The formul\ae~above for tensor perturbations are consistent with the approach of Ref.~\cite{Anselmi:2020lpp} in its range of validity (for $H<2M_2$, where $H$ is the Hubble rate at the inflationary scale).
By taking the ratio of~(\ref{Ptspectrum}) and  (\ref{power-spectrum}) one obtains the tensor-to-scalar ratio
\be r\equiv \frac{P_t}{P_R}= \frac{1}{1+\frac{2 H^2}{M_2^2}}\frac{8}{ \bar M_{\rm Pl}^2 N_{,i}N^{,i}}=\frac{r_E}{1+2H^2/M_2^2}. \label{rW}\ee

It is then clear that in order to quantify the effect of the massive spin-2 field on $r$ it is necessary to have $H$. This can be  obtained only after solving the dynamical system of scalar  equations (see Eqs.~(\ref{slow-roll-eq})). In Figs.~\ref{HLambda} and~\ref{HLambdaf} $H$ is given, respectively, as a function of $\rho$ and $f$  by requiring $P_R$ to match the observed value in~(\ref{PRPlanck}). Also the corresponding value of the symmetry-breaking scale parameter $\Lambda$ is provided in the same figures.  We notice that the observational  upper bound on $H$ in~(\ref{PlanckHboound}) is satisfied for all values of $\rho$ for $N=60$ in Fig.~\ref{HLambda} and for all values of $f$ for both $N=60$ and $N=55$ in Fig.~\ref{HLambdaf}. In Fig.~\ref{HLambda} we see a tension for $N=55$, but only when $\rho$ is small, that is in the pure-natural inflation limit.
We also checked that the observational bounds~\cite{Ade:2015lrj} on isocurvature perturbations are satisfied for all values of $\rho$ and $f$ considered in Figs.~\ref{HLambda} and~\ref{HLambdaf}.

\begin{figure}[t!]
\begin{center}
 \includegraphics[scale=0.50]{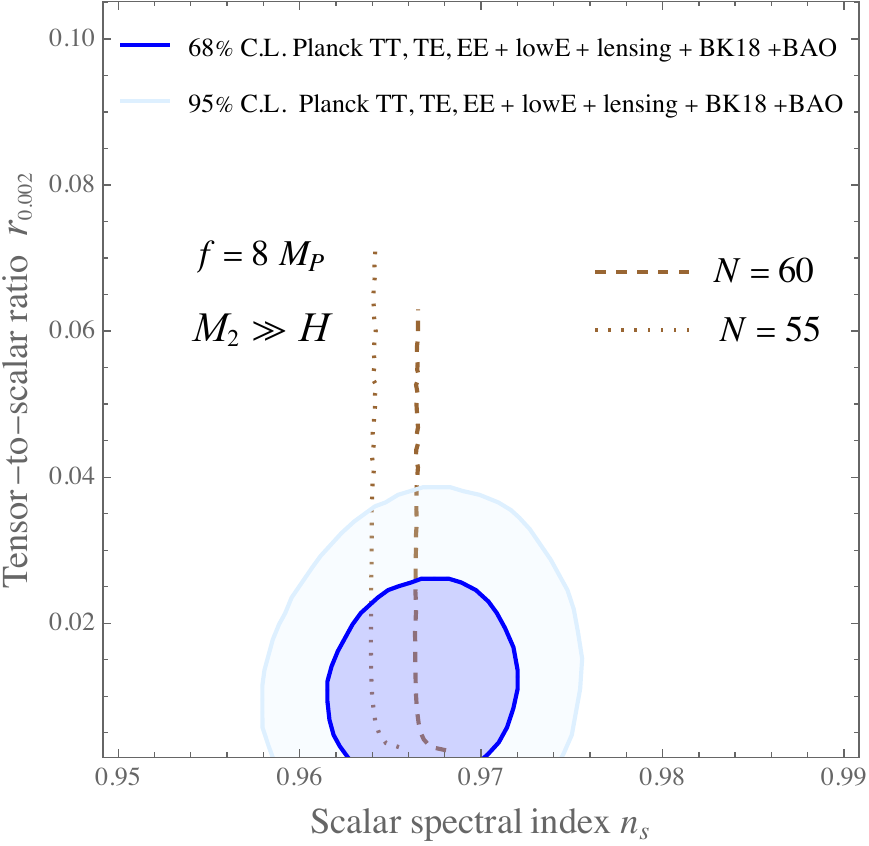}  
  \hspace{1.3cm}   \includegraphics[scale=0.50]{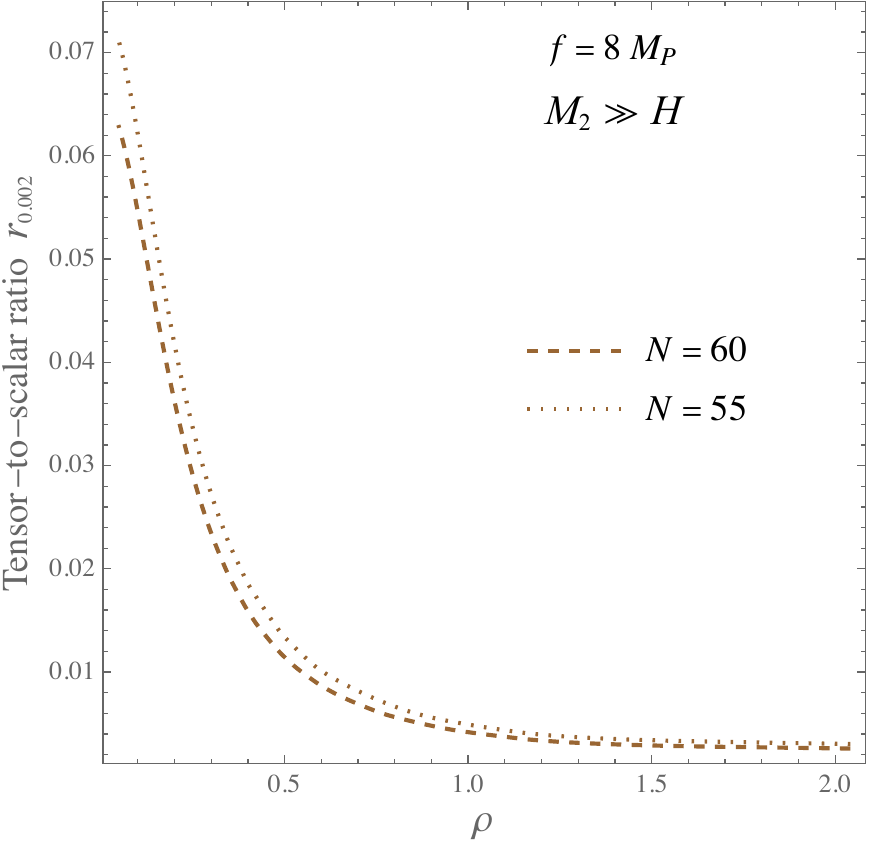} \\
    \vspace{1cm}
   \includegraphics[scale=0.50]{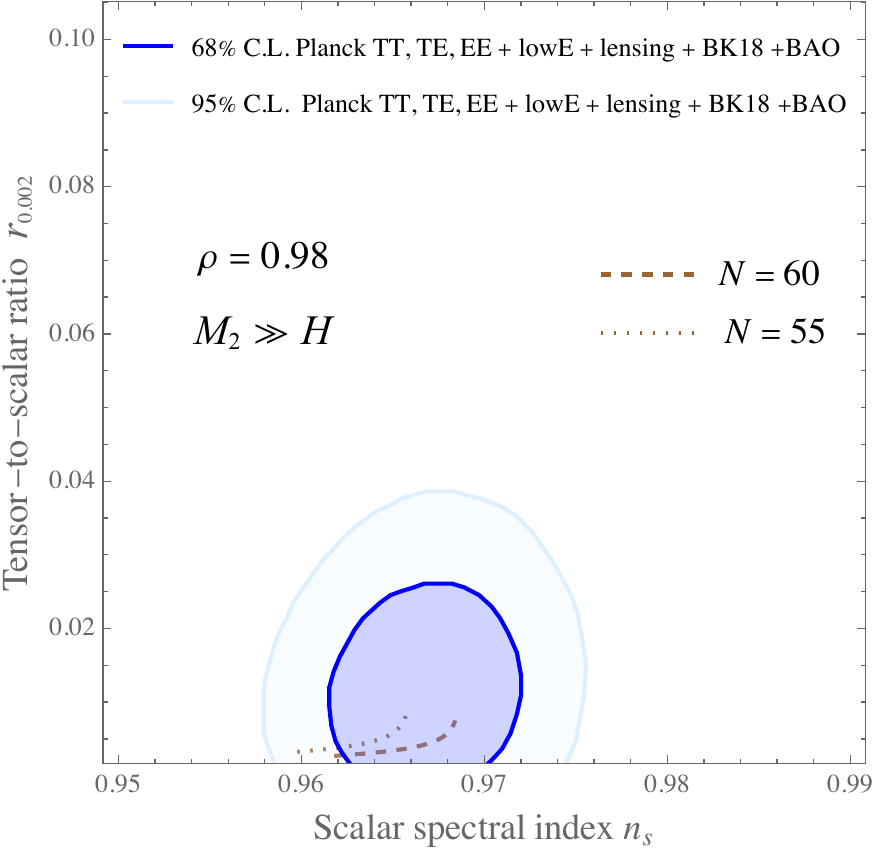}  
  \hspace{1.3cm}   \includegraphics[scale=0.50]{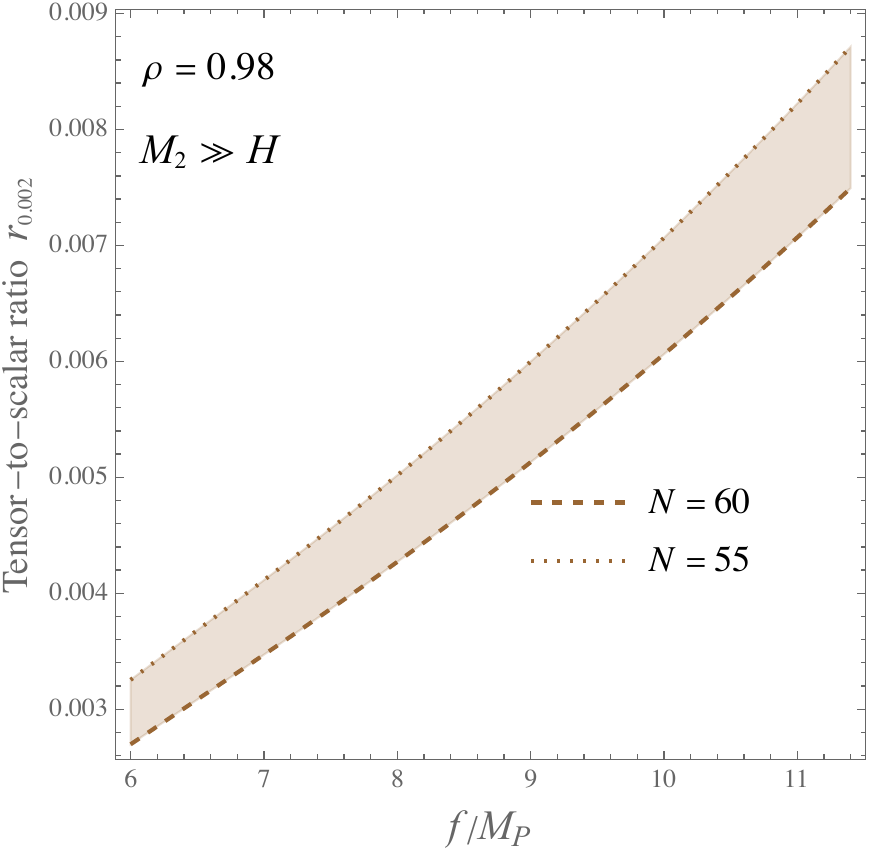} 
 	 
 \end{center}

   \caption{\em  {\bf Left plots:} the predictions for $n_s$ and $r$ compared with the observational constraints of Ref.~\cite{BICEP:2021xfz} and obtained by varying $\rho$ (upper plot) or $f$ (lower plot). {\bf Right plots:} the corresponding variation of the tensor-to-scalar ratio $r_{0.002}$.}\vspace{0.1cm}
   {\em

 Here $H$ is assumed to be negligibly small compared to the mass $M_2$ of the spin-2  field.
   }
\label{nsrNoGhost}
\end{figure}

\begin{figure}[h!]
\begin{center}
 \includegraphics[scale=0.50]{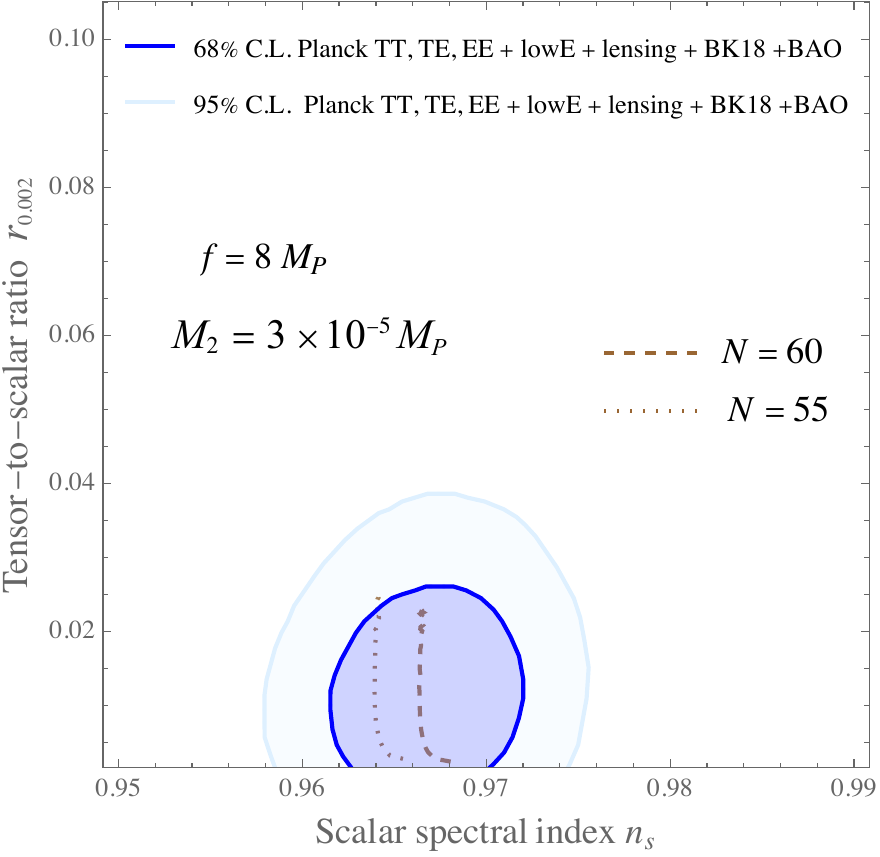}  
  \hspace{1.3cm}   \includegraphics[scale=0.50]{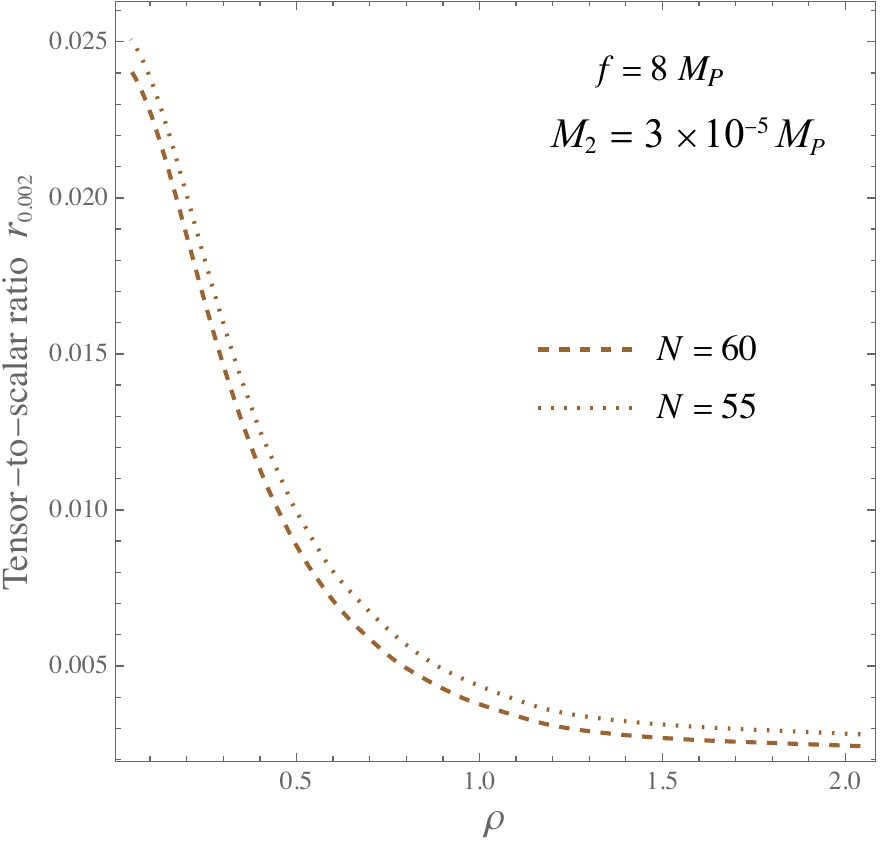} 
  \\
  \vspace{1cm}
  \includegraphics[scale=0.50]{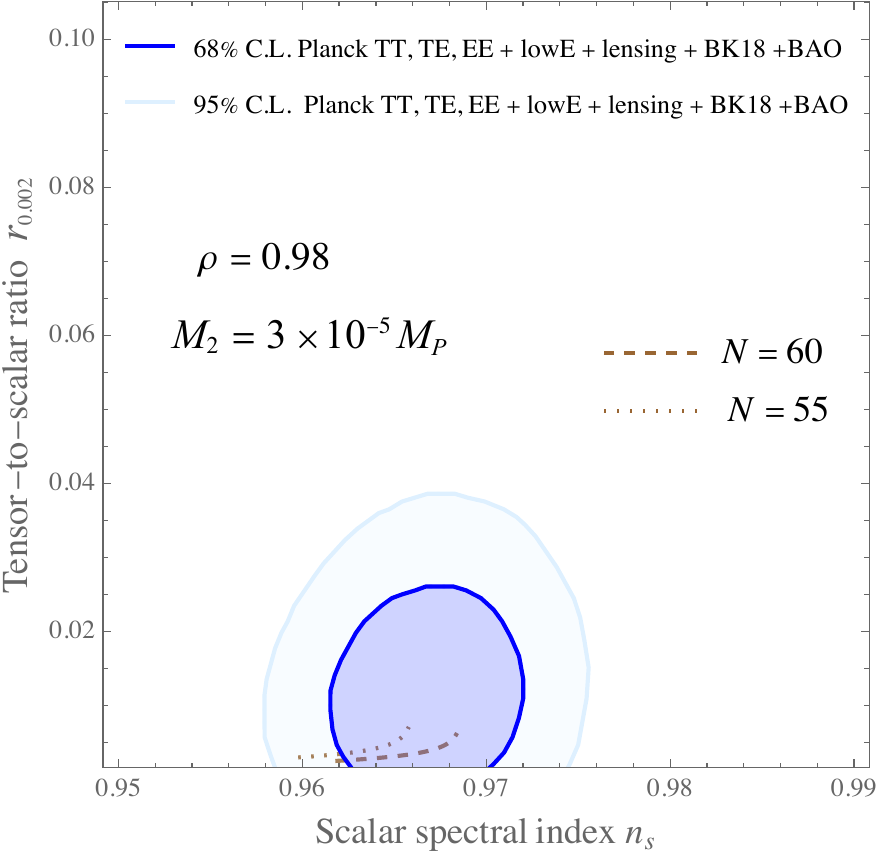}  
  \hspace{1.3cm}   \includegraphics[scale=0.50]{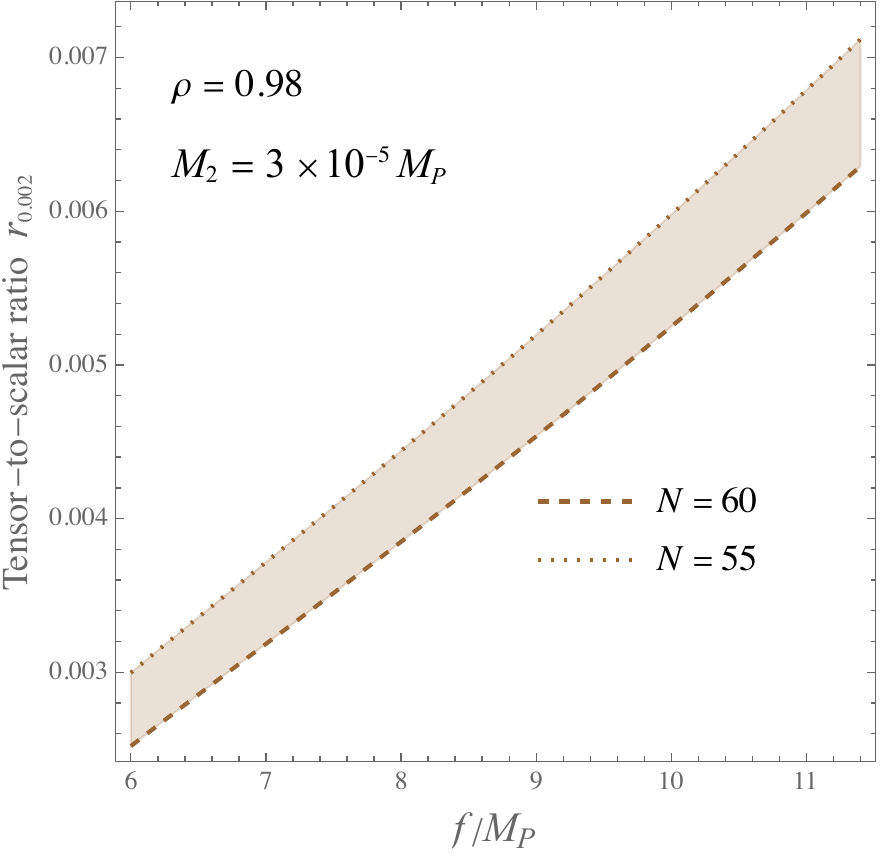} 
 	 
 \end{center}
   
   \caption{\em The same as in Fig.~\ref{nsrNoGhost}, but for a value of $M_2$ of order of $H$ at the inflationary scale (see Figs.~\ref{HLambda} and~\ref{HLambdaf}).} \label{nsrGhost}
\end{figure}

 The corresponding predictions for $n_s$ and $r$ are compared with the BK18 constrains in Fig.~\ref{nsrNoGhost} in the large $M_2/H$ limit (when the massive spin-2 field or ``fakeon" decouples) and in Fig.~\ref{nsrGhost} for $M_2=3\times 10^{-5}\bp \sim H$ (see Figs.~\ref{HLambda} and~\ref{HLambdaf}). 
 Note that this value of $M_2$ is compatible with the consistency condition $H < 2M_2$ of the fakeon approach in Ref.~\cite{Anselmi:2020lpp}  for all  values of $\rho$ and $f$, as clear from  Figs.~\ref{HLambda} and~\ref{HLambdaf}. Remarkably, for this   value (and all smaller values) of $M_2$  and $f\gtrsim 6\bp$ the predictions for  $n_s$ and $r$ satisfy the observational constraints, even when inflation is only driven by the PNGB\footnote{On the other hand, within the  EFT discussed in Sec.~\ref{CHIsecs}, the condition that $H$ should be well below $M_2$ precludes the compatibility between the theoretical prediction of $r$ and the observations in the pure natural-inflation case ($\rho\to 0$).}. The upper plots of Fig.~\ref{nsrNoGhost}, however, show that it is also possible to satisfy the constraints by taking $\rho$ large enough, i.e. by making  $z$ more active during inflation, even in the large $M_2/H$ limit. In that limit, as we have seen, the EFT discussed in Sec.~\ref{CHIsecs} holds.

 \section{Conclusions}\label{Conclusions}

Here we have studied the predictions of quadratic gravity in some of the best-motivated inflationary scenarios from the particle physics point of view that are ruled out by BK18 in Einstein gravity: CHI, which identifies the inflaton with the only elementary scalar in the SM and remains perturbative at the inflationary scales and above, and natural inflation (where the inflaton potential is naturally flat thanks to Goldstone's theorem). Higgs inflation and natural inflation have been combined here for the first time with quadratic gravity in its general form (featuring all quadratic-in-curvature terms in the action with dimensionless coefficients). Also, CHI has been realized in a model (the $a\nu$MSM) that allows to avoid any tension with the low-energy EW data, e.g. on $M_t$. In order to do so, the 1-loop RGEs of the relevant parameters in quadratic gravity coupled to the $a\nu$MSM have been presented. 

Quadratic gravity is a UV completion of Einstein's theory and its inflationary predictions have been worked out with two apparently different approaches in~\cite{Salvio:2017xul,Anselmi:2020lpp}. However, as shown in~\cite{Salvio:2020axm} and here for the first time in a generic multifield scenario, the two approaches lead to the same physical predictions in the regime of validity of the fakeon approach~\cite{Anselmi:2020lpp}, i.e. $H < 2M_2$, at least in the slow-roll approximation. This is true not only for the tensor and curvature power spectra, but also for the isocurvature modes due to possible further 4D scalars besides the inflaton. Moreover, we have shown that these physical predictions of the fakeon approach can also be reproduced in the effective field theory (the EFT) where the $W^2$ term is treated as a small perturbation and no further gravitational degrees of freedom are present besides the massless graviton. The EFT  holds, however, when $H$ is well below $M_2$.

Here it has been shown that both CHI and natural inflation can be made compatible with the strong BK18 constraints on $r$ and $n_s$ in quadratic gravity. The compatibility occurs in a large parameter space by taking $M_2$ small enough (not much bigger than $H$). Note that the agreement between the theoretical and experimental results in both models have been restored even for $H < 2M_2$ when the fakeon approach can be applied, and, in the natural-inflation case in the presence of the $R^2$ term, even when $H$ is well below $M_2$, which ensures the validity of the above-mentioned EFT.

%



\vspace{0.7cm}

 \footnotesize
\begin{multicols}{2}

\end{multicols}

\end{document}